\documentclass{article}
\usepackage{float}
\usepackage{spconf,amsmath,graphicx,cite}
\usepackage{comment}
\usepackage[a4paper, margin=2.5cm]{geometry}
\usepackage{algorithm}
\usepackage{algpseudocode}
\usepackage{amsmath}
\usepackage{bm}
\usepackage{multirow, multicol}
\usepackage{hyperref}
\usepackage{amsfonts}
\usepackage{xcolor}
\usepackage{footnote}
\usepackage{booktabs}
\usepackage{url}

\title{VRDMG: Vocal Restoration via Diffusion Posterior Sampling with Multiple Guidance}
\name{Author(s) Name(s)\thanks{Thanks to XYZ agency for funding.}}
\address{Author Affiliation(s)}

\name{\begin{tabular}{@{}c@{}}
Carlos Hernandez-Olivan$^{1,*}$\thanks{$^*$This work has been done during an internship at Sony} \qquad Koichi Saito$^2$ \qquad Naoki Murata$^2$ \qquad Chieh-Hsin Lai$^2$ \\ Marco A. Martínez-Ramírez$^2$ \qquad Wei-Hsiang Liao$^2$\qquad Yuki Mitsufuji$^{2,3}$
\end{tabular}}
\address{$^1$Dept. of Electronic Engineering and Communications, University of Zaragoza, Spain\\$^2$Sony AI, Tokyo, Japan \\ $^3$Sony Group Corporation, Tokyo, Japan}

\begin{document}
\ninept
\newcommand{\sigmaTilde}{\widetilde{\mkern-2mu\sigma}}
\newcommand{\xTilde}{\widetilde{\mkern-2mu x}}
\newcommand{\xHat}{\hat{x}}
\newcommand{\xPrime}{x'}
\newcommand{\wTilde}{\widetilde{\mkern-2mu w}}

\maketitle
\begin{abstract}
Restoring degraded music signals is essential to enhance audio quality for downstream music manipulation. 
Recent diffusion-based music restoration methods have demonstrated impressive performance, and among them, diffusion posterior sampling (DPS) stands out given its intrinsic properties, making it versatile across various restoration tasks.
In this paper, we identify that there are potential issues which will degrade current DPS-based methods' performance and introduce the way to mitigate the issues inspired by diverse diffusion guidance techniques including the RePaint (RP) strategy and the Pseudoinverse-Guided Diffusion Models ($\Pi$GDM). 
We demonstrate our methods for the vocal declipping and bandwidth extension tasks under various levels of distortion and cutoff frequency, respectively.
In both tasks, our methods outperform the current DPS-based music restoration benchmarks. 
We refer to \url{http://carlosholivan.github.io/demos/audio-restoration-2023.html} for examples of the restored audio samples.
\end{abstract}

\section{Introduction}
\label{sec:intro}

Music restoration is framed as an inverse problem with the objective of recovering a degraded music signal. In this domain, the tasks such as inpainting, declipping, bandwidth extension, and dereverberation are prevalent~\cite{narayanaswamy21_interspeech}. 
These tasks are ill-posed, indicating that multiple restoration outcomes could be equally valid.
Various techniques have been proposed to address these issues, ranging from unsupervised methods based on signal processing to supervised deep neural network (DNN) approaches using paired degraded and clean signals~\cite{Záviška2021IEEEJSTSP, Imort2022ISMIR, Kuleshov2017ICLR, Moliner2023IEEETASLP}. 
These methods may not operate robustly when the target audio deviates from the underlying assumptions or the characteristics of the paired data used for training. Additionally, we need to prepare assumptions and paired data for each task.

On the other hand, recent approaches utilize deep generative models trained solely on clean signals as a prior to retrieve the clean signal from observed data~\cite{survey,cqtdiff,10095761,narayanaswamy21_interspeech,turetzky22_interspeech,murata2023gibbsddrm}. 
This paradigm does not require degraded signals during training, allowing for a data-driven acquisition of assumptions solely about the clean signals. 
It offers the advantage of being adaptable to various tasks without the need for retraining. Among these, notably, unsupervised diffusion-based approaches have demonstrated significant performance \cite{survey,cqtdiff,10095761,murata2023gibbsddrm}. 

CQT-Diff\cite{cqtdiff} addresses challenges in piano declipping, bandwidth extension, and inpainting by employing diffusion posterior sampling (DPS)\cite{dps} in conjunction with the reconstruction guidance (RG)~\cite{Ho2022NIPS} and the data consistency (DC) method~\cite{Choi2021ICCV}.
While the method demonstrates impressive performance, there remains room for improvement in several aspects.
Expanding from RG, the Pseudoinverse-Guided Diffusion Models ($\Pi$GDM)~\cite{pigdm} have been introduced to address inverse problems not only rooted in noisy and non-linear degraded measurements but also when confronted with a non-differentiable degradation function.
Moreover, empirical evaluations by the authors reveal $\Pi$GDM outperform DPS with even fewer diffusion steps, indicative of its computational efficiency.
Although the DC method has been recognized for its proficient performance, it is not without its limitations. 
RePaint (RP)~\cite{repaint} highlighted that the application of the DC method for image inpainting results in the predicted region mirroring the texture of the adjacent regions, yet being semantically inconsistent. 
The implications of semantic inaccuracy introduced by the DC method potentially degrading performance in music restoration tasks.
Furthermore, while the empirical validation of CQT-Diff has been confined to piano data, its efficacy on datasets with diverse distributions, such as vocal data\cite{Virtanen2008CombiningPI}, remains unexplored.

In this paper, we enhance DPS-based music restoration and evaluate its efficacy on diverse vocal data. 
Our contribution can be summarized as five parts.
We introduce the RP strategy~\cite{repaint} with time scheduling inspired by FreeDoM~\cite{freedom}. Additionally, we refine RG by incorporating time-dependent scaling and the $\Pi$GDM method. 
Given the inherent orthogonality of these approaches, we conduct comprehensive experiments on declipping and bandwidth extension task and ascertain the most effective combination.
Comparing inference speeds between the CQT-Diff (CQT model) and SaShiMi-Diff (1D model) architectures from \cite{cqtdiff}, by applying our techniques, we find the 1D model achieves comparable performance to the CQT model but boasts an 15x faster inference time.


\begin{figure*}
    \centering
    \includegraphics[width=15cm]{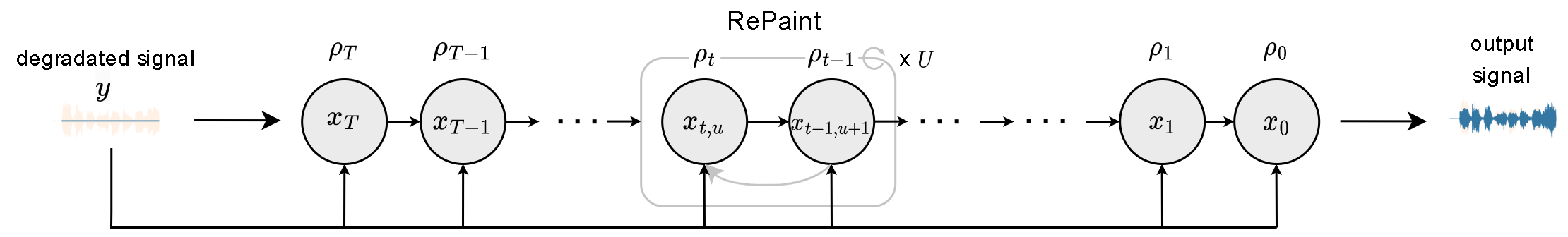}
    \vspace{-10pt}
    \caption{Inference process of our RG $\delta\rho$ + DC + RP method. $\rho_{t}$ is the parameter that controls the gradients weighting and $\bm{x}_{t}$ is the prediction at each diffusion step $t$. $U$ are the repainting steps.}
    \vspace{-10pt}
    \label{fig:scheme}
\end{figure*}

\section{Related work}
\label{sec:related_work}

Prior to the application of diffusion models to audio restoration, problem-specific models were introduced for challenges like declipping \cite{Záviška2021IEEEJSTSP,Imort2022ISMIR} and bandwidth extension \cite{Moliner2023IEEETASLP, 10095382}. 
Following the success of unsupervised diffusion-based approaches in image domain, there has been a growing interest in problem-agnostic frameworks\cite{dps,KawarEES22,murata2023gibbsddrm,pigdm}. 
Music restoration spans various challenges, all potentially benefiting from unified approaches adept at addressing these distinct tasks via posterior sampling. 
Several unsupervised diffusion-based methods with this aim have emerged, focusing on piano data~\cite{cqtdiff} or vocal data~\cite{10095761}. 
CQT-Diff~\cite{cqtdiff} offers a unified strategy for piano restoration, effectively recovering piano signals for a range of tasks and their specific degradations: inpainting, bandwidth extension, and declipping. 

To the best of our knowledge, DPS \cite{dps} and its variants, which have found applications in image domain~\cite{pigdm,repaint,freedom}, remain inadequately explored for audio restoration tasks. In this study, we showcase the potential enhancements achievable by transitioning techniques from image domain, such as $\Pi$GDM~\cite{pigdm} and RePaint \cite{repaint}, to the realm of audio domain, thereby amplifying the efficacy of RG and DC.

\section{Vocal Restoration via Diffusion Posterior sampling}
\label{sec:model}

\subsection{Score-based Generative Modeling}
Let us denote the data distribution of clean signals by $p_{\rm{data}}$.
Consider a diffusion process wherein a clean signal $\bm{x}_{0} \sim p_{\rm{data}}$ is progressively diffused into an intermediate noisy signal $\bm{x}_{t}$ over time step $t=0, \cdots, T$ with associated noise levels $\sigma_{T} = \sigma_{\rm{max}} > \cdots > \sigma_{t} > \sigma_{t-1} > \cdots > \sigma_{0} = \sigma_{\rm{min}}$. 
The terminal point $\bm{x}_{T}$ is characterized as Gaussian noise $\bm{x}_{T} \sim \mathcal{N}(0, \sigma^{2}_{\rm{max}}\bm{I})$.

Diffusion models \cite{sohl2015deep,ddpm,edm}, alternatively termed score-based models, generate a clean signal through the reverse diffusion process. These models approximate the gradient of the logarithmic probability density relative to the score function $\nabla_{\bm{x}_{t}}\log p_{t}(\bm{x}_{t})$ \cite{Hyvarinen2005JMLR}.
The reverse diffusion process can be articulated by the following Stochastic Differential Equation (SDE)~\cite{edm}:
\begin{equation}
      \mathrm{d}\bm{x}_{t} = \left(\frac{1}{2}g(t)^{2} - \dot{\sigma(t)}\sigma(t) \right)\nabla_{\bm{x}_{t}}\log p_{t}(\bm{x}_{t}) \mathrm{d}t + g(t) \mathrm{d}\bm{w}_{t}
\label{eq:sde}
\end{equation}
where $w$ denotes the standard Wiener process, $\dot{\sigma}(t)$ represents the time derivative of noise levels at time $t$, and $g(t)$ is \textit{diffusion} coefficient of $\bm{x}_{t}$ \cite{Song2021ICLR}.

In the diffusion models, the score function is approximated utilizing a neural network $D_{\bm{\theta}}(\bm{x}; \sigma)$, where $\bm{\theta}$ are the parameters of neural network. 
In EDM~\cite{edm}, upon which our work is predicated, the score function is estimated as $\nabla_{\bm{x}_{t}}\log p_{t}(\bm{x}_{t}; \sigma_{t}) = (D_{\bm{\theta}}(\bm{x}_{t}; \sigma_{t}) - \bm{x}_{t})/ \sigma_{t}^{2}$, where the neural network of the denoiser $D_{\bm{\theta}}(\bm{x}_{t}; \sigma_{t})$ is trained with the following $L_2$ objective:
\begin{equation}
      \mathbb{E}_{\bm{x}_{\rm{data}} \sim p_{\rm{data}}, \bm{n} \sim \mathcal{N}(0, \sigma_{t}^{2}\bm{I})} \left[\| D_{\bm{\theta}}(\bm{x}_{\rm{data}} + \bm{n}; \sigma_{t}) - \bm{x}_{\rm{data}}\|^{2}_{2} \right], 
    \label{eq:objective}
\end{equation}
where $\bm{x}_{\rm{data}} \sim p_{\rm{data}}$ is a training data.
By solving the SDE by substituting the estimated score function with discrete sampling times $t$, the models can generate samples.

\subsection{Inverse Problem via Posterior Sampling}

The objective of music inverse problem is to recover the clean music signal $\bm{x}_{0}$ from observed degraded signal $\bm{y} = \mathcal{A}(\bm{x}_{0}) + \bm{\epsilon}$, where $\mathcal{A}(\cdot)$ represents the degradation function, $\bm{\epsilon} \sim \mathcal{N}(0, \sigma^{2}_{\bm{y}}I)$ is the measurement noise and $\sigma_{\bm{y}}^{2}$ is noise variance of measurement.
To address inverse problems through diffusion models, we can substitute the score $\nabla_{\bm{x}_{t}}\log p_{t}(\bm{x}_{t})$ in Eq. \ref{eq:sde} for the conditional score fucntion $\nabla_{\bm{x}_{t}}{\log}p_{t}(\bm{x}_{t}|\bm{y})$ \cite{dps,cqtdiff}.
Within the Bayesian framework, the \textit{posterior} $p(\bm{x}|\bm{y})$ can be expressed as $p(\bm{x}|\bm{y}) \propto p(\bm{x})p(\bm{y}|\bm{x})$.
Leveraging this relationship, the conditional score $\nabla_{\bm{x}_{t}}{\log}p_{t}(\bm{x}_{t}|\bm{y})$ can be deduced as:
\begin{equation}
    \nabla_{\bm{x}_{t}}{\log}p_{t}(\bm{x}_{t}|\bm{y}) = \nabla_{\bm{x}_{t}}{\log}p_{t}(\bm{x}_{t}) + \nabla_{\bm{x}_{t}}{\log}p_{t}(\bm{y}|\bm{x}_{t}).
    \label{eq:dps}
\end{equation}
In Eq.~\eqref{eq:dps}, it's essential to evaluate two components; the first term is the score function which we can compute via $(D_{\bm{\theta}}(\bm{x}_{t}; \sigma_{t}) - \bm{x}_{t})/ \sigma_{t}^{2}$ and the second term is likelihood which can be computed as\cite{dps}:
\begin{equation}
 \nabla_{\bm{x}_{t}}{\log}p_{t}(\bm{y}|\bm{x}_{t}) \simeq -\rho(t)\nabla_{\bm{x}_{t}}\|\bm{y}- \mathcal{A}(D_{\bm{\theta}}(\bm{x}_{t}; \sigma_{t}))\|^{2}.
 \label{eq:guidance}
\end{equation}
We refer to this method reconstruction guidance (RG)~\cite{cqtdiff,Ho2022NIPS}.

The gradient in Eq. \ref{eq:guidance} is scaled by a time-dependent scaling function $\rho(t)$.
In CQT-Diff~\cite{cqtdiff}, $\rho(t)$ is empirically determined so that $\rho(t) = \rho\prime \sqrt{L} /(t\|G\|^{2})$, where $G = \nabla_{\bm{x}_{t}}\left\| \bm{y}-\mathcal{A}(D_{\bm{\theta}}(\bm{x}_{t}; \sigma_{t})) \right\|^2$, where $L$ is the sample length of audio signal and $\rho\prime$ is a constant value.
In our work, drawing inspiration from the segmented stages of diffusion sampling as depicted in \cite[Fig.3]{freedom}, we empirically find that redefining $\rho(t)$ as $\rho(t) \delta\rho$, where $\delta\rho = (T - t) / 75$ exhibits superior performance compared to a static $\rho\prime$. 
Defining $\delta\rho$ in this manner, the scale of RG remains small during the initial stages of the diffusion process (or the chaotic phase) where the generation is out of control according to FreeDoM \cite{freedom}. 
As the diffusion process progresses and the generation becomes more controllable, the scale of RG becomes large.

\subsection{Pseudo-Inverse Guidance}
The second term of Eq.~\eqref{eq:guidance} can alternatively be calculated by utilizing Pseudo-inverse Guidance ($\Pi$GDM) \cite{pigdm} as:
\begin{equation} \label{eq:pigdm}
    \nabla_{\bm{x}_{t}}{\log}p_{t}(\bm{y}|\bm{x}_{t}) \simeq \left((h^{\dagger}(\bm{y}) - h^{\dagger}(h(\hat{\bm{x}}_{0})))^{\top}\frac{\partial\hat{\bm{x}}_{0}}{\partial\bm{x}_{t}} \right)^{\top},
\end{equation}
where $\hat{\bm{x}}_{0}$ is estimated clean signal by the denoiser network $D_{\bm{\theta}}(\bm{x}_{t}; \sigma_{t})$, $h$ is a non-linear degradation function, and $h^{\dagger}$ need to adhere to the equation $h(h^\dagger(h(\bm{x}))) = h(\bm{x})$ for all $\bm
{x}$. $h$ and $h^{\dagger}$ can be even non-differentiable degradation function, rendering DPS inapplicable.
Empirical evaluations by the authors \cite{pigdm} reveal $\Pi$GDM outperform RG with even fewer diffusion steps, indicative of its computational efficiency.

\subsection{Improving Data Consistency method}
In previous study \cite{cqtdiff}, the DC method is applied during sampling.
The DC method substitutes the intermediate prediction $\hat{\bm{x}}_{0} = D_{\bm{\theta}}(\bm{x}_{t}; \sigma_{t})$ with the measurement $\bm{y}$ every discrete sampling step.
On the image inpainting task, RePaint \cite{repaint} points out that only applying the DC method leads that the predicted region only matches the texture of the adjacent regions, thereby being semantically inaccurate.
This phenomenon can potentially be observed in inverse problem utilizing the DC method on music domain, leading to worse restoration performance.

To alleviate this limitation, we advocate integrating the RePaint (RP) strategy~\cite{repaint} with the DC method.
The RP strategy is a resampling strategy that first diffuses output $\bm{x}_{t-1}$ back to $\bm{x}_{t}$ through the sampling as:
\begin{equation} \label{eq:repaint}
    \bm{x}_{t} \sim \mathcal{N}(\bm{x}_{t-1}, (\sigma_{t}^{2}-\sigma_{t-1}^{2}) \textbf{I})
\end{equation}
and then conducts regular reverse diffusion step from $\bm{x}_{t}$, again.
We designate this one cycle as the RP cycle.
Originally, the RP strategy is applied with every diffusion step during inference with multiple the RP cycle (Please refer to \cite[Algo.~1]{repaint}).
Even in the absence of the RP strategy, diffusion models generally need multiple sampling steps to generate samples, which is time consuming.
Hence, applying multiple RP cycles will result in an increased number of sampling steps.
To circumvent this issue, we propose to apply the RP cycle exclusively during the phase which crucially affects sampling outcome inspired by the categorized stages of diffusion sampling ~\cite[Fig.3]{freedom}.
We define the number of the RP cycles $U$ depending on the entire number of the diffusion steps $T$ as:
\begin{equation} \label{eq:repaint_phase}
    U = \left\{ \begin{array}{lcc}
             0 &   \forall  & t < \phi_{1} \ \textit{T} / 3 
             \\ \textit{u} &   \forall  & \phi_{1} \ \textit{T} / 3 \le
 t \le \phi_{2} \ \textit{T} / 3 
             \\ 0 &   \forall  &   t > \phi_{2} \ \textit{T} / 3
             \end{array}
   \right.
\end{equation}
where $\phi_{1}$ and $\phi_{2}$ are term to define when the RP cycle starts and end in inspired by FreeDoM~\cite{freedom}.

\section{Experiments and Results}\label{sec:results}

\subsection{Dataset}
\label{sec:data}
To evaluate the methods under the various types of vocals, we use vocal data which do not contain audio effects (dry vocals) and contain audio effects (wet vocals).
For pre-training of diffusion models, we use the NHSS dataset \cite{nhss}, which only contains dry vocals, and an internal dataset of wet vocals. 
The NHSS dataset contains 100 English pop songs (20 unique songs) of ten different male and female singers.
The total duration of the dataset is 95.76h for training and 0.94h for validation. 
For evaluation, we use the NUS dataset \cite{nus}, which contains dry vocals, and the wet vocals from MUSDB18 dataset \cite{musdb18}.
We dowmsample all the tracks to 22.05~kHz. 
Due to the time constraints, We selected 200 random segments of 2.97 seconds for both dry and wet vocals, ensuring that they encompassed a diverse range of vocalists.

\subsection{Model Configurations}
In our preliminary experiments, we observe that the CQT-Diff (CQT model) requires more time to output restored data compared to the SaShiMi-Diff (1D model), due to its inverse CQT implementation \cite{cqtdiff,Velasco2011DAFx}. Both models are referred in \cite[Section 3.1]{cqtdiff}.
Consequently, we train those two types of models and evaluate both in terms of performance and inference time.
The difference between these two models is the input data representation for the neural network.
In the 1D model, the neural network receives input in the form of a waveform representation. 
On the other hand, in the CQT model, the input data representation is CQT representation.

Both models have a U-Net architecture comprising the six layers with the dilated convolutions and the GELU activations. 
Both models have 15M parameters.
We use Adam optimizer and an initial learning rate of $2e-4$. 
The noise variance schedule at diffusion step $t$, $\sigma_{t}$, follows the expression:
\begin{equation}
    \sigma_{t} = \left( \sigma_{\rm{max}}^{\frac{1}{\nu}} + \frac{t}{T-1}\left(\sigma_{\rm{min}}^{\frac{1}{\nu}} - \sigma_{\rm{max}}^{\frac{1}{\nu}}\right)\right)^\nu \quad t = T \ldots 1
\end{equation}
We set $\nu = 13$, $\sigma_{\rm{min}} = 10^{-4}$, $\sigma_{\rm{max}} = 1$ and $S_{\rm{churn}}=5$ for the noise variance schedule \cite{edm}. 
For inference step, we utilize $T=300$ diffusion steps. 

\subsection{Declipping}

\begin{table*}[ht]
    \centering
    \scriptsize
    \caption{Objective metrics for declipping with pre-trained CQT and 1D models. For each metric and test set, each result is highlighted in bold and second best result is underlined.}
    \begin{tabular}{
    c| p{2.5cm}| c c c c | c c c c
    }
        \hline
        & & \multicolumn{4}{c|}{SDR = 5dB} & \multicolumn{4}{c}{SDR = 10dB} \\
        & & \multicolumn{2}{c}{FAD} & \multicolumn{2}{c|}{SI-SDR} & \multicolumn{2}{c}{FAD} & \multicolumn{2}{c}{SI-SDR}\\
         & Method & $\downarrow$ dry & $\downarrow$ wet & $\uparrow$ dry & $\uparrow$ wet & $\downarrow$ dry & $\downarrow$ wet & $\uparrow$ dry & $\uparrow$ wet \\
        \cmidrule{1-10}
        & Clipped & 3.48 & 2.25 & 6.42 & 5.06 & 1.10 & 0.87 & 11.08 & 9.18\\
        \cmidrule{1-10}
        \multirow{5}{*}{1D} & RG \cite{cqtdiff} & 1.84 & 1.55 & 9.79 & 5.36 & 0.66 & 1.19 & 14.39 & 7.76 \\
        & RG + DC & 0.94 & 1.05 & 10.13 & \underline{6.15} & 0.66 & 0.52 & 14.48 & \underline{8.69} \\
        & RG $\delta \rho$ + DC & \underline{0.88} & \underline{1.00} & \underline{10.97} & 6.03 & \underline{0.35} & \underline{0.32} & \underline{14.79} & 8.57 \\
        &$\Pi$GDM + DC & 2.76 & 1.73 & 5.44 & 3.18 & 1.02 & 0.76 & 10.31 & 7.01\\
        &RG $\delta \rho$ + DC + RP & $\mathbf{0.47}$ & $\mathbf{0.65}$ & $\mathbf{11.83}$ & $\mathbf{7.03}$ & $\mathbf{0.15}$ & $\mathbf{0.21}$ & $\mathbf{15.42}$ & $\mathbf{9.05}$\\
        \cmidrule{1-10}
        \multirow{5}{*}{CQT} & RG \cite{cqtdiff} & 1.59 & 1.05 & 9.87 & 5.76 & 0.37 & 1.04 & 14.29 & 8.20 \\
        &RG + DC & 1.16 & \underline{0.80} & 10.09 & 6.24 & 0.37 & 0.39 & 14.33 & \underline{8.91} \\
        &RG $\delta \rho$ + DC & $\mathbf{0.52}$ & $\mathbf{0.34}$ & \underline{10.58} & \underline{6.51} & \underline{0.16} & $\mathbf{0.19}$ & \underline{14.50} & $\mathbf{9.02}$ \\
        &$\Pi$GDM + DC & 2.62 & 1.57 & 5.70 & 3.75 & 0.74 & 0.60 & 10.70 & 7.22 \\
        &RG $\delta \rho$ + DC + RP & \underline{0.73} & 1.02 & $\mathbf{11.61}$ & $\mathbf{7.33}$ & $\mathbf{0.13}$ & \underline{0.28} & $\mathbf{15.03}$ & 8.75\\
        \hline
    \end{tabular}
    \label{tab:declipping}
\end{table*}

\begin{table*}[!h]
    \centering
    \scriptsize
    \caption{Objective metrics for bandwidth extension with pre-trained CQT and 1D models}
    \begin{tabular}{
    c| p{2.5cm}| c c c c | c c c c
    }
        \cmidrule{1-10}
        & & \multicolumn{4}{c|}{$f_c$ = 3KHz} & \multicolumn{4}{c}{$f_c$ = 5KHz} \\
        & & \multicolumn{2}{c}{FAD} & \multicolumn{2}{c|}{LSD} & \multicolumn{2}{c}{FAD} & \multicolumn{2}{c}{LSD} \\
        \cmidrule{3-10}
        Model & Method & $\downarrow$ dry & $\downarrow$ wet & $\downarrow$ dry & $\downarrow$ wet & $\downarrow$ dry & $\downarrow$ wet & $\downarrow$ dry & $\downarrow$ wet \\
        \cmidrule{1-10}
        & LPF & 4.12 & 3.16 & 4.26 & 4.55 & 2.53 & 1.79 & 3.61 & 3.86\\
        \cmidrule{1-10}
        \multirow{5}{*}{1D} & RG \cite{cqtdiff} & 2.38 & \underline{1.35} & $\mathbf{1.87}$ & 1.95 & 1.72 & 1.00 & \underline{1.59} & 1.86\\
        & RG + DC \cite{cqtdiff} & 1.53 & 1.44 & \underline{1.98} & $\mathbf{1.84}$ & \underline{0.57} & $\mathbf{0.44}$ & 1.67 & \underline{1.61}\\
        & RG $\delta \rho$ + DC & $\mathbf{1.13}$ & 1.38 & 2.01 & 1.89 & $\mathbf{0.46}$ & \underline{0.48} & $\mathbf{1.57}$ & \underline{1.61}\\
        &$\Pi$GDM + DC & 1.69 & $\mathbf{1.26}$ & 2.14 & \underline{1.86} & 0.67 & \underline{0.48} & 1.93 & 1.70\\
        &RG $\delta \rho$ + DC + RP & \underline{1.15} & 1.46 & 2.00 & 1.88 & 0.73 & 0.58 & 1.56 & $\mathbf{1.60}$\\
        \cmidrule{1-10}
        \multirow{5}{*}{CQT} & RG \cite{cqtdiff} & \underline{1.63} & 0.75 & $\mathbf{1.95}$ & 1.91 & 1.31 & 0.82 & 1.70 & 1.79\\
        &RG + DC \cite{cqtdiff} & $\mathbf{1.06}$ & \underline{0.65} & \underline{2.01} & \underline{1.82} & $\mathbf{0.39}$ & $\mathbf{0.30}$ & \underline{1.67} & $\mathbf{1.59}$\\
        &RG $\delta \rho$ + DC & 1.65 & 1.17 & 2.34 & 2.10 & \underline{0.44} & 0.35 & $\mathbf{1.64}$ & \underline{1.61}\\
        &$\Pi$GDM + DC & $\mathbf{1.06}$ & $\mathbf{0.63}$ & 2.00 & $\mathbf{1.80}$ & 1.23 & \underline{0.32} & 1.68 & 1.66 \\
        &RG $\delta \rho$ + DC + RP & 1.78 & 1.57 & 2.31 & 2.03 & 0.51 & 0.67 & 1.68 & 1.62\\
        \hline
    \end{tabular}
    \label{tab:band}
\end{table*}

Declipping an audio signal is restoring the signal after a distortion with a threshold $c$ has been applied to it. The degradation function is formulated as $\mathcal{A}(\bm{x}_{0})=(\lvert \bm{x}_{0} + c \rvert - \lvert \bm{x}_{0} - c \rvert) / 2$. 
This is hard-clipping setting~\cite{Imort2022ISMIR}. 
We perform both objective and subjective evaluations. 
The objective evaluation is measured in terms of the Frèchet Audio Distance (FAD)~\cite{fad} and the Scale-Invariant Signal-to-Distortion Ratio (SI-SDR)~\cite{Roux2019ICASSP}. 

We test both models with the different sampling methods: RG and RG + DC as the baseline methods proposed in \cite{cqtdiff}, and our proposed methods: RG $\delta \rho$ + DC, which consists on applying RG with DC method and $\delta \rho$, RG $\delta \rho$ + DC + RP and DC + $\Pi$GDM. 
With employing the RP strategy, Eq.~\ref{eq:repaint} is invoked with $u=10$, $\phi_{1}=1.5$ and $\phi_{2}=2.8$, leading to the $U=10$ RP cycles between the diffusion steps $t=150$ to $t=20$.
For the $\Pi$GDM method, we define $h(\bm{x}) = \mathcal{A}(\bm{x})$ and $h^{\dagger}(\bm{x}) = \mathcal{A}(\bm{x})$.
With this definition, $h(h^\dagger(h(\bm{x}))) = h(\bm{x})$ is satisfied, as illustrated in Eq.~\ref{eq:pigdm}.
We compare the models and methods with two Signal-to-Distortion Ratio (SDR) of applied clipping , 5dB and 10dB as it is reported in Table \ref{tab:declipping}.


In terms of the 1D model, RG $\delta \rho$ + DC + RP consistently outperforms all of the other methods for all the experimental settings.
As for the CQT model, upon overall consideration, RG $\delta \rho$ + DC + RP tends to show better performance than the other methods.
From these observations, combining all the method, which are $\delta \rho$, the DC method, and the RP strategy, is effective to enhance the performance of the original RG method.
Moreover, comparing with RG $\delta \rho$ + DC + RP of the 1D and the CQT model, it is remarkable that the 1D model tends to show better scores since the inference time of the 1D model is $\times 15$ faster than the CQT model using the a RTX 3090 24Gb GPU.
The actual inference time of the 1D model using RG $\delta \rho$ + DC + RP is three minutes and the CQT model takes 45 minutes.

\begin{figure}[!t]
    \centering
    \includegraphics[width=\columnwidth]{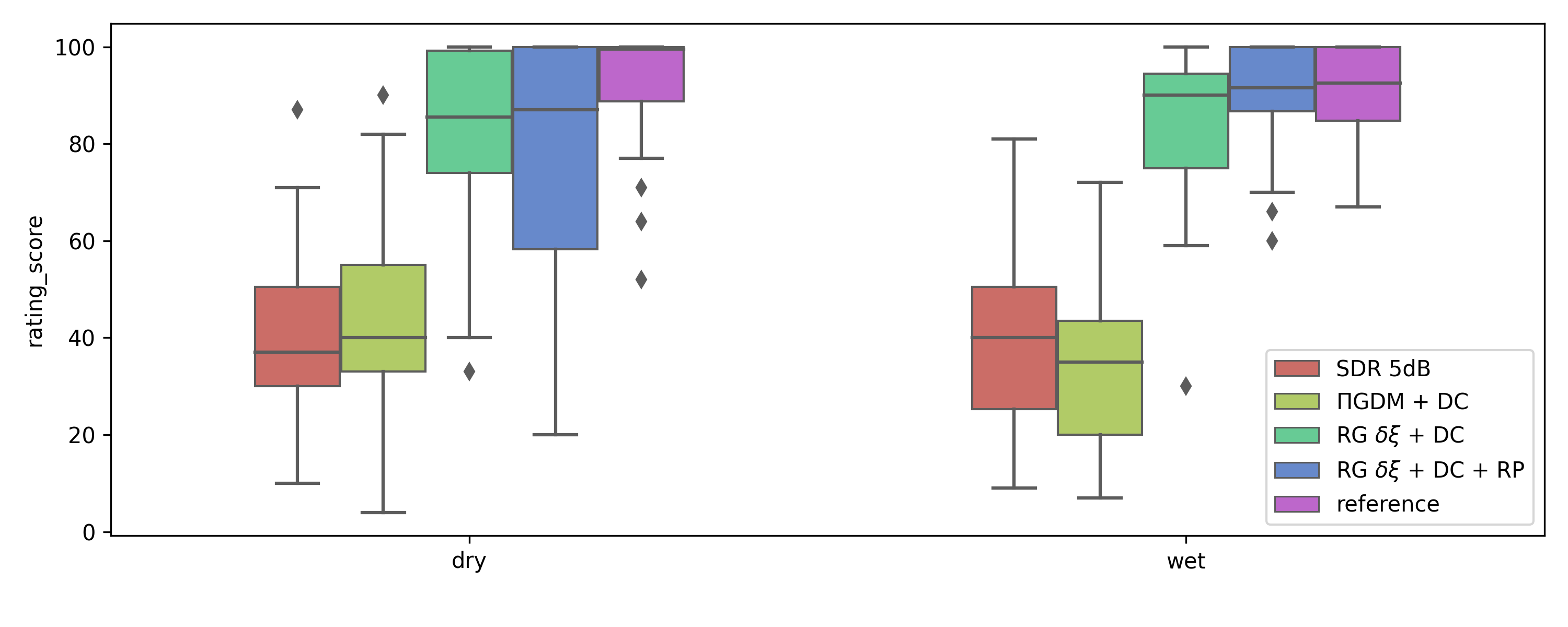}
    \vspace{-12pt}
    \caption{Results of the audio declipping listening test of CQT model for 5dB.}
    \vspace{-12pt}
    \label{fig:declip_mushra}
\end{figure}

We also conducted a listening test using the multiple stimuli with the hidden reference and anchor (MUSHRA) protocol \cite{ITU-R-BS.1534-3}. 
We ask the 12 participants to rate the quality of the restored signals with the different methods.
To avoid the listening fatigue, we only include the samples from the CQT models under SDR = 5dB, which is harsher setting as the objective evaluation proves. 
Also, we discarded the worst-performing methods (RG) and (RG + DC) according to the objective results in Table \ref{tab:declipping}. 
The results of the listening test are illustrated in Fig \ref{fig:declip_mushra}.
In both dry and wet configurations, RG $\delta \rho$ + DC + RP show the best score as we see in Table \ref{tab:declipping}. 
Notably, within the wet setting, RG $\delta \rho$ + DC + RP is perceived by the listeners to be very close to the reference.

\subsection{Bandwidth Extension}
Bandwidth extension consists on restoring the frequencies that have been removed from the signal with a low pass filter (LPF).
Within this configuration, the degradation function of bandwidth extension is formulated as $\mathcal{A}(\bm{x}) = \mathrm{LPF}(\bm{x})$. 
We evaluate the methods with two settings of cutoff frequencies $f_{c}=3$ and $5$~kHz.
For the $\Pi$GDM method, same as the declipping setting, we define $h(\bm{x}) = \mathcal{A}(\bm{\bm{x}})$ and $h^{\dagger}(\bm{x}) = \mathcal{A}(\bm{x})$.
With this definition, if we assume that the LPF can cutoff all the frequency components lower than $f_{c}$, $h(h^\dagger(h(\bm{x}))) = h(\bm{x})$ is satisfied, as illustrated in Eq.~\ref{eq:pigdm} since applying consecutive the LPF with the same $f_c$ to the signal is the same as applying it once. In this task, we set $\phi_1=2.5$, $\phi_2=2.8$ and $u=5$ for the RP strategy.

\begin{figure}
    \centering
    \includegraphics[width=\columnwidth]{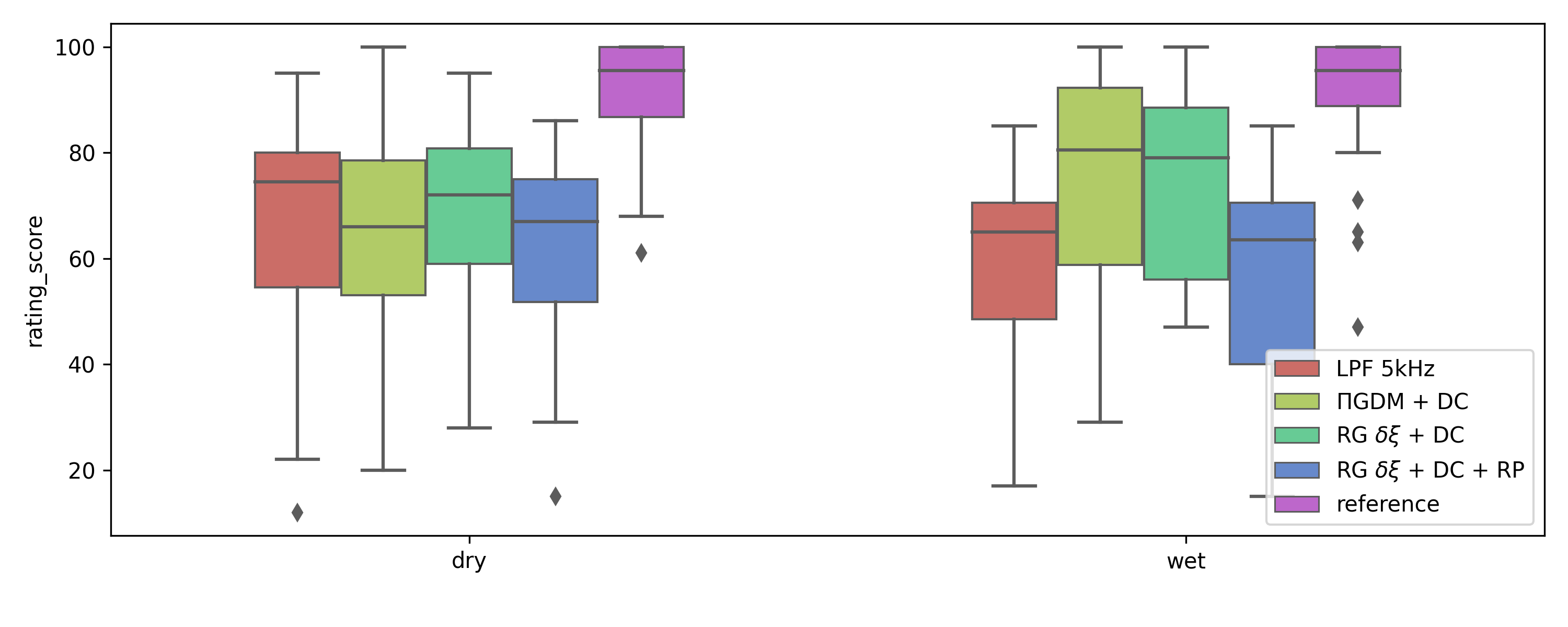}
    \vspace{-12pt}
    \caption{Results of the audio bandwidth extension listening test of CQT model for 5kHz.}
    \vspace{-12pt}
    \label{fig:band_mushra}
\end{figure}

In this case, the objective evaluation shows that the best-performing methods are RG + DC and $\Pi$GDM + DC. 
In contrast with the declipping task, $\Pi$GDM performs well and outperforms other techniques like RG. 
Is also worth noticing that RP does not work that well in this task. 
The reason why RP does not work for this task might be due to the diffusion trajectories that the time-traveling introduces itself which for bandwidth extension are not optimal.

In parallel to the declipping task, we conducted a listening test, engaging the same participants to evaluate the methods for the $f_c=5$~kHz case. 
The results of the bandwidth extension listening test are presented in Fig. \ref{fig:band_mushra}. 
The subjective evaluation reports that the best method for $f_{c}=5$~kHz is RG $\delta \rho$ + DC.
The listening test results show that any of the methods outperform the lowpassed signal for the dry vocals. 
The human auditory system is highly attuned to the formants of speech. Consequently, due to the absence of effects in dry signals containing only the voice, the inadequate reconstruction of these formants might be more perceptible to the human ear compared to wet signals where effects mask this imperfect formant reconstruction.
In wet signals, we can observe how $\Pi$GDM and RG $\delta \rho$ + DC perform better, which also happens in the objective evaluation.

\section{Conclusions}

In this study, we analyzed various diffusion sampling techniques for music restoration. We pointed out potential shortcomings in the current DPS-based methods and introduced the way to address them.
By incorporating the RP strategy, $\Pi$GDM, and time-dependent weights for the guidance term, we enhanced restoration performance in both vocal declipping and bandwidth extension tasks, outclassing the baseline methods. Notably, the 1D model, when using the same sampling technique, achieved equivalent performance to the CQT model but was 15 times faster.

%
%
%

\bibliographystyle{IEEEbib}
\bibliography{refs}

\end{document}